\documentclass[prl,aps,amssymb,twocolumn,showpacs,floatfix]{revtex4}
\usepackage[dvips]{graphicx}
\def\be{\begin{equation}}
\def\ee{\end{equation}}
\def\bea{\begin{eqnarray}}
\def\eea{\end{eqnarray}}

\begin{document}
 
\title{Transport of charge-density waves in the presence of disorder:\\
Classical pinning vs quantum localization}
\author{A.D.~Mirlin$^{1,2,*}$}
\author{D.G.~Polyakov$^{1,\dagger}$}
\author{V.M.~Vinokur$^{3}$}
\affiliation{$^{1}$Institut f\"ur Nanotechnologie,
Forschungszentrum Karlsruhe, 76021 Karlsruhe, Germany \\
$^{2}$Institut f\"ur Theorie der kondensierten Materie, Universit\"at
Karlsruhe, 76128 Karlsruhe, Germany\\
$^{3}$Materials Science Division, Argonne National Laboratory, Argonne, 
IL 60439, USA}
 
\date{\today}
\begin{abstract} We consider the interplay of the elastic pinning and the
Anderson localization in the transport properties of a charge-density wave in
one dimension, within the framework of the Luttinger model in the limit of
strong repulsion. We address a conceptually important issue of which of the
two disorder-induced phenomena limits the mobility more effectively. We argue
that the interplay of the classical and quantum effects in transport of a very
rigid charge-density wave is quite nontrivial: the quantum localization sets
in at a temperature much smaller than the pinning temperature, whereas the
quantum localization length is much smaller than the pinning length.
\end{abstract}
 
\pacs{71.10.Pm, 73.21.-b, 73.63.-b, 73.20.Jc}

\maketitle

A confluence of ideas formulated for mesoscopic disordered electron systems on
one side and for strongly correlated clean systems on the other has brought
forth a new field---mesoscopics of strongly correlated electron
systems. Prominent examples of such systems include charge-density waves
(CDWs), Wigner crystals, and Luttinger liquids. Low-energy excitations in
these systems are of essentially collective nature and are described in terms
of elastic waves. In the presence of static disorder, the conductivity of the
systems is strongly suppressed at low temperatures, which is commonly referred
to as pinning, or localization.

A key concept in the mesoscopics of disordered electron systems is that of the
Anderson localization \cite{anderson58}. This phenomenon is due to the quantum
interference of multiply-scattered electron waves on spatial scales larger
than the localization length $\xi_{\rm loc}$. The localization is destroyed by
inelastic electron-electron (e-e) scattering on the scale of the dephasing
length. The notions of weak localization and dephasing due to e-e scattering,
established for Fermi-liquid systems \cite{altshuler85}, have recently been
shown to be also applicable to Luttinger liquids \cite{gornyi05a}.

On the other hand, considering the interplay of disorder and interaction in
the opposite limit of a very strong Coulomb interaction, one arrives at the
concept of {\it pinning} of elastic waves on the spatial scale of the pinning
length $\xi_{\rm pin}$ \cite{larkin70}. This concept has a long history
\cite{feigelman89,blatter94} and applies not only to CDWs and similar electron
systems but, more generally, to every elastic object in a random environment,
ranging from domain walls in ferromagnets and ferroelectrics to vortex
lattices in type-II superconductors. However, as far as strongly correlated
electron systems are concerned, the notions of Anderson localization and
pinning are often viewed in the literature as essentially
synonymous. Specifically, the Anderson localization of electrons and the
pinning of CDWs are thought of as two sides of the same phenomenon which
gradually evolves with changing strength of e-e interaction.

However, it is useful to recall that the physics of localization and pinning,
as we know them from the pioneering works by Anderson \cite{anderson58} and
Larkin \cite{larkin70}, respectively, is distinctly different: the
localization is a quantum phenomenon, whereas the pinning is essentially
classical. This distinction has important consequences. In particular, in
classical elastic systems at vanishing coupling to the external thermal bath,
any inelastic scattering results in the activated temperature behavior of the
mobility at low $T$. On the other hand, in the case of the Anderson
localization, it was argued that
in the limit of vanishing coupling to the bath a disordered system of weakly
interacting one- or two-dimensional electrons cannot support either activated
or variable-range hopping transport at low $T$, undergoing instead a
metal-insulator transition at a {\it finite} critical temperature $T_c$
\cite{gornyi05b}.

With this background in mind, we now formulate the main question we are
concerned with here: {\it What is the relation between 
pinning and localization?} Which of them restricts
the mobility of strongly correlated CDWs more effectively? Specifically, we
aim to understand what are the ratios of the characteristic scales in energy
space and in real space, $T_c/\Delta_{\rm cl}$ and $\xi_{\rm loc}/\xi_{\rm
pin}$, respectively, where $\Delta_{\rm cl}$ is the pinning temperature
\cite{remark1} [defined below in Eq.~(\ref{6})]. In this Letter, we focus on
transport of CDWs in one dimension (1d) in the limit of strong interaction,
i.e., on the case of very rigid CDWs, and consider spinless electrons.

Let us specify the model. The spatial modulation of the charge density
$\rho(x)$ is written in terms of a smoothly varying phase $\phi(x)$ of a
single-harmonic CDW as $\rho=-\partial_x\phi/\pi+
\cos\,[2(k_Fx-\phi)\,]/\pi\lambda$. The Hamiltonian $H=H_{\rm el}+H_{\rm
kin}+H_{\rm dis}$ is given by a sum of the elastic part 
$H_{\rm el}=c\int \!dx \,(\partial_x\phi)^2/2$, the
kinetic part $H_{\rm kin}=\pi v_F\int\!dx\,\Pi^2/2$,
%\be 
%H_{\rm el}={c\over 2}\int \!dx \,(\partial_x\phi)^2~\,,\quad H_{\rm
%kin}={\pi v_F\over 2}\int\!dx\,\Pi^2~,
%\label{2}
%\ee
where $\Pi$ is the momentum conjugated to $\phi$ (throughout the paper
$\hbar=1$), and the part $H_{\rm dis}$ describing backscattering of electrons
off a static random potential:
\be
H_{\rm dis}={1\over 2\pi\lambda}\int\!dx\,\left[\,V_b(x)e^{-2i\phi(x)}
+{\rm h.c.}\,\right]~.
\label{3}
\ee
The disorder is taken to be of white-noise type with the correlators
$\overline{V_b(x)V_b^*(0)}=w\delta(x)$ and $\overline{V_b(x)V_b(0)}=0$.  The
elastic constant $c$ and the Fermi velocity $v_F$ are input
parameters of the low-energy theory and include Fermi-liquid-type
renormalizations coming from the ultraviolet scales. The electron current $j$
is related to $\Pi$ via the Fermi velocity: $j=v_F\Pi$, so that the velocity
of elastic waves is given by $u=(\pi v_Fc)^{1/2}$. The system can
thus be characterized by two velocities, $c$ (up to $\hbar$, the elastic
constant in 1d is velocity) and $u$, the former describing the static
properties of a CDW and the latter its dynamics. Their ratio
\be
K=u/\pi c\ll 1
\label{4}
\ee
is known as the Luttinger constant, and is the main parameter of our
consideration.

The classical limit corresponds to $K\to 0$ with the rigidity $c={\rm
const}$. To see the significance of this limit, consider the correlator of the
phase $\phi$ in the Matsubara representation ${\cal D}(x,\tau)
=\left<\phi(0,0)[\phi(0,0)-\phi(x,\tau)]\right>$, which in the absence of
disorder is given by
\bea
{\cal D}_0(x,\tau)={K\over 4}\ln\left[\,\left({u\over \pi
T\lambda}\right)^2\sinh{\pi Tx_+\over
u}\sinh{\pi Tx_-\over u}\,\right],
\label{5}
\eea
where $x_\pm=x\pm iu\tau$ and a proper ultraviolet cutoff is assumed, e.g.,
$x\!\to\! x +\lambda\,{\rm sgn}\,x$. With decreasing $K$ the system rapidly
acquires rigidity: at zero $T$, the long-range order is only broken on large
length scales at $\ln (|x|/\lambda)\sim 1/K$. In the classical limit, 
quantum fluctuations of $\phi$ vanish and only
thermal fluctuations with ${\cal D}_0(x,\tau)=T|x|/2c$ are present. At the same
time, the classical limit is also the static limit ($H_{\rm kin}\to 0$).
In the presence of disorder, the long-range order is broken in the classical
limit already at $T=0$, so that ${\cal D}(x,\tau)$, averaged over
disorder, becomes a function of $x/\xi_{\rm pin}$. The pinning length
$\xi_{\rm pin}$ and the amplitude $\Delta_{\rm cl}$ of elastic-energy
fluctuations on the scale of $\xi_{\rm pin}$ 
(assuming that $\xi_{\rm pin}\gg\lambda$) are
\be
\xi_{\rm pin}=\lambda\left(c^2/w\lambda\right)^{1/3}~,\quad \Delta_{\rm
cl}=c/\xi_{\rm pin}~.
\label{6}
\ee 
%where it is assumed that $\xi_{\rm pin}\gg\lambda$. 

Slightly beyond the classical limit, at $0<K\ll 1$, quantum fluctuations
soften the pinning potential by changing the exponent of the pinning length
$\xi_{\rm pin}\to \lambda (c^2/w\lambda)^{1/(3-2K)}$. This effect is of little
importance in our discussion [and of no importance whatsoever if $K\ll
1/\ln(c^2/w\lambda)]$. The essential quantum effects are the onset of dynamics
in the system, with a characteristic velocity $u$, and the emergence of the
{\it quantum localization}. The latter is characterized by the localization
length $\xi_{\rm loc}$ and the ``localization temperature" $T_1$, below which
the localization effects become strong \cite{recall}. We will see below that
the behavior of $\xi_{\rm loc}$ and $T_1$ as a function of $K$ for $K\ll 1$ is
highly nontrivial.

%To recall the results for weak interaction ($a=1-K\ll 1$), it was recently
%shown \cite{gornyi05a} that for spinless electrons in the Luttinger liquid
%$T_1\sim (w/a^2u)(u^2/w\lambda)^{2a/(1+2a)}$ and the critical temperature
%$T_c\sim aT_1/\ln(1/a)\ll T_1$. The localization length $\xi_{\rm loc}$
%behaves for $a^2T_1\alt T\alt T_1$ as $\xi_{\rm loc}\sim
%(u^2/w)(T\lambda/u)^{2a}$. These results cannot be straightforwardly
%generalized to the case of strong interaction with $K\ll 1$, in particular
%because the velocities $u$ and $c$ become then parametrically different from
%each other.  %\cite{remark3}.

In this Letter, we rely on the conventional bosonization (poorly suited to
study the localization and dephasing but conveniently treating strong and weak
interaction in the disordered system with essentially the same effort). We
explore the large-$\omega$ expansion of the ac conductivity to extract the
relevant parameters of the system \cite{feigelman81}.

The conductivity is given by the Kubo
formula \cite{giamarchi04} with the current $j=i\partial_\tau \phi/\pi$\,:
%\bea
%\sigma (\omega,T)&=&-{1\over i\omega}\,{e^2v_F\over \pi}
%-{1\over i\omega}\,{e^2\over
%\pi^2}\left\{\int_0^{1/T}\!\!d\tau \,e^{i\Omega_n\tau}\int\! dx\,
%\overline{\left<T_\tau \dot{\phi}(x,\tau)\dot{\phi}(0,0) 
%\right>}\right\}_{i\Omega_n\to\omega+i0}~. 
%\label{7}
%\eea
\bea
\sigma (\omega,T)&=&-{1\over i\omega}\,{e^2v_F\over \pi}
-{1\over i\omega}\,{e^2\over
\pi^2}\left\{\int_0^{1/T}\!\!d\tau \,e^{i\Omega_n\tau}\right.\nonumber\\
&\times&\left.\int\! dx\,
\overline{\left<T_\tau \dot{\phi}(x,\tau)\dot{\phi}(0,0) 
\right>}\right\}_{i\Omega_n\to\omega+i0}~. 
\label{7}
\eea
%The replicated action $S=S_0-S_{\rm dis}$ for the averaging over
%$\phi$ in Eq.~(\ref{7}) reads
The averaging over $\phi$ in Eq.~(\ref{7}) is performed with the replicated
action $S=S_0-S_{\rm dis}$, where
\bea
\hspace{-4mm}S_0&=&{c\over 2}\sum_m\int \!dx\!\int \!d\tau  
\left[ (\partial_x\phi_m)^2 +
{1\over u^2}\,(\partial_\tau\phi_m)^2\right]~,
\label{8}\\
\hspace{-4mm}S_{\rm dis}&=&{w\over (2\pi\lambda)^2}\sum_{m,m'}
\int \!dx\!\int \!d\tau\!\int
\!d\tau'\nonumber\\
&\times&\cos \left[\,2\phi_m(x,\tau)-2\phi_{m'}(x,\tau')\,\right]~.
\label{9}
\eea
%\bea
%\hspace{-4mm}S_0&=&{c\over 2}\sum_m\int \!dx\!\int \!d\tau  
%\left[ (\partial_x\phi_m)^2 +
%{1\over u^2}\,(\partial_\tau\phi_m)^2\right]~,
%\label{8}\\
%\hspace{-4mm}S_{\rm dis}&=&{w\over (2\pi\lambda)^2}\sum_{m,m'}
%\int \!dx\!\int \!d\tau\!\int
%\!d\tau'\cos \left[\,2\phi_m(x,\tau)-2\phi_{m'}(x,\tau')\,\right]~.
%\label{9}
%\eea
Inspection of Eqs.~(\ref{7})-(\ref{9}) shows that $\sigma(\omega, T)$ is
expandable in powers of disorder as
\bea
\sigma (\omega,T)&=&-{e^2 v_F\over \pi}\,{1\over
i(\omega+i0)}\sum_{N=0}\left(
{\Delta_{\rm cl}\over T}\right)^{3N}\!\!
\left({\pi T\lambda\over u}\right)^{2KN}
\!\!\nonumber\\
&\times&f_N\!\left(\Omega_n/T\right)\vert_{i\Omega_n\to\omega+i0}~,
\label{10}
\eea
%\bea
%\sigma (\omega,T)&=&-{e^2 v_F\over \pi}\,{1\over
%i(\omega+i0)}\sum_{N=0}\left(
%\Delta_{\rm cl}/T\right)^{3N}\!\!
%\left(\pi T\lambda/u\right)^{2KN}
%f_N\!\left(\Omega_n/T\right)\vert_{i\Omega_n\to\omega+i0}~,
%\label{10}
%\eea
which is at the same time the large-$\omega$ expansion. The functions $f_N$
are dimensionless (parametrized by the constant $K$) real functions of
$\Omega_n/T$, with $f_0=1$.

At first order in $w$ we have a contribution to $\sigma (\omega,T)$ in terms
of ${\cal D}_0(x,\tau)$ [Eq.~(\ref{5})]:
\bea
&&\sigma_1 (\omega,T)=-{4\over i\omega}{e^2\over \pi^2}{w\over
(2\pi\lambda)^2}\nonumber \\
&&\times\left[\,D_n^2\,(2C_0-C_n-C_{-n})\,\right]_{i\Omega_n\to\omega+i0}~,
\label{11}
\eea
where 
\bea
D_{n\neq 0}&=&\int\! d\tau \,e^{i\Omega_n\tau}\!\int \!dx 
\,\partial_\tau{\cal D}_0(x,\tau)={i\pi
v_F\over \Omega_n}~,\label{12}\\
C_n&=&\int\!d\tau\,e^{i\Omega_n\tau-4{\cal D}_0(0,\tau)}~,
\label{13}
\eea
%\bea
%&&\sigma_1 (\omega,T)=-{4\over i\omega}{e^2\over \pi^2}{w\over
%(2\pi\lambda)^2}
%\left[\,D_n^2\,(2C_0-C_n-C_{-n})\,\right]_{i\Omega_n\to\omega+i0}~,\\
%\label{11}
%\eea
%where $D_n$ and $C_n$ are expressed through the propagator 
%${\cal D}_0(x,\tau)$ [Eq.~(\ref{5})] as
%\bea
%&&D_{n\neq 0}=\int\! d\tau \,e^{i\Omega_n\tau}\!\int \!dx 
%\,\partial_\tau{\cal D}_0(x,\tau)={i\pi
%v_F\over \Omega_n}~,\quad
%C_n=\int\!d\tau\,e^{i\Omega_n\tau-4{\cal D}_0(0,\tau)}~,
%\label{13}
%\eea
$D_0=0$. When calculating
the difference $2C_0-C_n-C_{-n}$ in Eq.~(\ref{11}), the ultraviolet cutoff in
Eq.~(\ref{5}) can be neglected for $K<3/2$ and $\sigma_1(\omega,T)$ is then
explicitly obtained as
\bea 
&&\sigma_1 (\omega,T)={2ie^2v_F\Delta_{\rm cl}^3\over \pi T\omega^3}\,
\left({2\pi T\lambda\over u}\right)^{2K}
\!K^2\Gamma(1-2K)\nonumber\\ &&\times\left[\,{1\over
\Gamma^2(1-K)}-{\sin \pi K\over \pi}\, {\Gamma(K-i{\omega\over 2\pi T})\over
\Gamma(1-K-i{\omega\over 2\pi T})}\,\right]~.
\label{14}
\eea
%\bea 
%\sigma_1 (\omega,T)&=&-{2e^2v_F\over i\pi\omega}\,{\Delta_{\rm cl}^3\over
%T\omega^2}\left({\pi T\lambda\over u}\right)^{2K}\nonumber \\
%&\times&2^{2K}K^2\Gamma(1-2K)\left[\,{1\over
%\Gamma^2(1-K)}\right.\nonumber\\&-&\left.{\sin \pi K\over \pi}\,
%{\Gamma(K-i{\omega\over 2\pi T})\over
%\Gamma(1-K-i{\omega\over 2\pi T})}\,\right]~.  
%\label{14}
%\eea
%\bea 
%&&\sigma_1 (\omega,T)=-{2e^2v_F\over i\pi\omega}\,{\Delta_{\rm cl}^3\over
%T\omega^2}\left({2\pi T\lambda\over u}\right)^{2K}
%\!K^2\Gamma(1-2K)\nonumber\\ &&\times\left[\,{1\over
%\Gamma^2(1-K)}-{\sin \pi K\over \pi}\, {\Gamma(K-i{\omega\over 2\pi T})\over
%\Gamma(1-K-i{\omega\over 2\pi T})}\,\right]~.
%\label{14}
%\eea
%%%\bea 
%%%&&\sigma_1 (\omega,T)=
%%%\left(2ie^2v_F\Delta_{\rm cl}^3/\pi T\omega^3\right)
%%%\left(2\pi T\lambda/u\right)^{2K}
%%%\!K^2\Gamma(1-2K)\nonumber\\ &&\times\left[\,
%%%\Gamma^{-2}(1-K)-\pi^{-1}\sin (\pi K)\, \Gamma(K-i\omega/2\pi T)\,
%%%\Gamma^{-1}(1-K-i\omega/2\pi T)\,\right]~.
%%%\label{14}
%%%\eea
An important observation is that in Eq.~(\ref{14}) there appears a
characteristic scale $\omega\sim KT$.  Specifically, for $K\ll 1$ the real
part of $\sigma_1 (\omega,T)$ is written as
%\bea
%{\rm Re}\,\sigma_1 (\omega,T)&\simeq& {4e^2v_F\over
%\pi^3\omega}\left({\omega_{\rm pin}\over \omega}\right)^3
%\eta(\omega,T)
%{\omega^2\,\over
%\omega^2+(2\pi KT)^2}~,
%\label{15}
%\eea
\bea
{\rm Re}\,\sigma_1 (\omega,T)\simeq 4e^2v_F\omega_{\rm pin}^3\eta/
\pi^3\omega^2
[\omega^2+(2\pi KT)^2]~,
\label{15}
\eea
where $\omega_{\rm pin}=u/\xi_{\rm pin}=\pi K\Delta_{\rm cl}$ is the pinning
frequency and $\eta(\omega,T)=(\lambda\max\{\omega,T\}/u)^{2K}$ is a weak
function which describes the renormalization of disorder by quantum
fluctuations [can be neglected for $K\ll
1/\ln(u/\lambda\max\{\omega,T\})$]. 
The scaling of ${\rm Re}\,\sigma_1$ in Eq.~(\ref{15}) with
the ratio $\omega/KT$ reflects the difference in the energy
scales characterizing the behavior of $\sigma (\omega,T)$ at zero $T$ and that
of the dc conductivity $\sigma_{\rm dc}$ as a function of $T$. The
characteristic $\omega$ for the zero-$T$ ac conductivity is $\omega_{\rm
pin}$, whereas the characteristic $T$ for $\sigma_{\rm dc}$ is $\Delta_{\rm
cl}\sim \omega_{\rm pin}/K$.

The ac conductivity can be equivalently rewritten as
$\sigma(\omega,T)=e^2v_F/\pi[-i\omega+M(\omega,T)]$, where the
current-relaxation rate ${\rm Re}M$ depends, in general, on $\omega$ and
$T$. For high $T\gg\max\{\Delta_{\rm cl},T_1\}$, when both pinning and
localization are suppressed, the $\omega$ dispersion of $\sigma (\omega,T)$
should obey the Drude formula, i.e., all terms in $M$ of order in $w$
higher than one can be neglected. The first-order expansion (\ref{15}) allows
us to extract the Drude-relaxation rate in the CDW state (here and below we
omit the factor $\eta$):
\be
{\rm Re}\,M(\omega,T)=4\omega_{\rm pin}^3/\pi^2
\left[\,\omega^2+(2\pi KT)^2\,\right]~.
\label{16}
\ee 
The dc conductivity for $T\gg \max\{\Delta_{\rm cl},T_1\}$ thus reads
\be
\sigma_{\rm dc}(T)=\pi^3 K^2(e^2 v_F/\omega_{\rm pin}^3)T^2=\pi K
(e^2\lambda^2/w)T^2~.
\label{17}
\ee
The corresponding backscattering time at $\omega=0$ is
\be
\tau (T)=\omega_{\rm pin}^{-1}\left(\pi T/\Delta_{\rm cl}\right)^2~.
\label{18}
\ee

The dependence of $\tau$ on $K$ for small $K$ is of central importance to
us. First of all, let us note that, although $\sigma_{\rm dc}$ decreases with
increasing strength of interaction, $\tau$ diverges in the classical limit
$K\to 0$, $c={\rm const}$. This divergency suggests that the quantum
localization is destroyed at temperature $T_1$ which is much smaller than the
classical scale $\Delta_{\rm cl}$.

To find $T_1$, we use the general condition, valid in 1d for arbitrary
strength of interaction, 
\be
\tau (T_1)\sim \tau_\phi(T_1)~,
\label{19}
\ee
where $\tau_\phi(T)$ is the phase-breaking time \cite{remark5}. For
$\tau_\phi\ll\tau$, the quantum-interference effects are suppressed on the
ballistic scale and so reduce to the weak-localization correction to the
conductivity (see Ref.~\onlinecite{gornyi05a} for a derivation of this
correction at $1-K\ll 1$), whereas for $\tau_\phi\gg\tau$ the localization,
which develops on the scale of the mean free path, is only slightly affected
by the inelastic processes. Within the bosonization framework, one of the
regular ways to extract $\tau_\phi$ is to proceed to the third order $(N=3)$
in Eq.~(\ref{10}), which is a minimal order that exhibits the Anderson
localization \cite{remark2}. In this paper, however, we do not follow this
intricate path and rely on a heuristic argument leading in the limit of strong
interaction to
\be
\tau_\phi^{-1}(T_1)\sim T_1~.
\label{20}
\ee

For a weakly interacting Luttinger liquid \cite{gornyi05a},
$\tau_\phi^{-1}(T)\sim (1-K)[T/\tau(T)]^{1/2}$, which gives
$\tau^{-1}_\phi(T_1)\sim (1-K)^2T_1$. It is important that (i) soft inelastic
scattering with energy transfers $\omega'\ll \tau_\phi^{-1}(T)$ is not
effective in dephasing the localization effects (similarly to 
Fermi liquids \cite{altshuler82}) and (ii) characteristic $\omega'\alt
T$. Hence, the upper bound for $\tau_\phi^{-1}(T)$ is the temperature
itself. Piecing together these results [and assuming that $\tau_\phi^{-1}(T)$
is a monotonic function of the interaction strength], we arrive at
Eq.~(\ref{20}). Combining Eqs.~(\ref{18})-(\ref{20}) we get
\be
T_1\sim K^{1/3}\Delta_{\rm cl}~,
\label{21}
\ee
i.e., the quantum localization sets in when $\sigma_{\rm dc}$ is already
strongly suppressed by the pinning.

To find $\xi_{\rm loc}$, we first note that the mean free path $l(T)$ at
$T\ll\Delta_{\rm cl}$ turns out to be much smaller than $\xi_{\rm pin}$. As
follows from Eq.~(\ref{18}), for excitations characterized by velocity $u$:
$l(T)=u\tau(T)=\xi_{\rm pin}(\pi T/\Delta_{\rm cl})^2$. Since in 1d the
localization is established on the scale of the mean free path, we have for
the localization length $\xi_{\rm loc}(T)$ at temperature $T_1$:
\be
\xi_{\rm loc}\sim K^{2/3}\xi_{\rm pin}~.
\label{22}
\ee
This is the same spatial scale at which the Giamarchi-Schulz
\cite{giamarchi88,giamarchi04} renormalization-group flow enters the
strong-coupling regime \cite{maslov95}. Note that it is the localization that
develops at this scale, not pinning; which is worth emphasizing in view of the
Anderson localization length and the pinning length being conventionally
thought to be the same in CDWs.

We thus see that both $T_1/\Delta_{\rm cl}$ and $\xi_{\rm loc}/\xi_{\rm pin}$
are small in CDWs, which is quite remarkable since the relative strength of
pinning and localization depends on whether we consider scaling in temperature
or in real space. In the ultraviolet limit of large $T$ or small spatial scale
$L$ neither pinning nor localization is important. With decreasing $T$, the
classical pinning ``wins", i.e., happens at $T\sim\Delta_{\rm cl}\gg T_1$. A
subtle point, however, is that with increasing $L$, it is the quantum
localization that wins, i.e., it develops at $L\sim\xi_{\rm loc}\ll \xi_{\rm
pin}$. This unusual behavior reflects the peculiarity of the very rigid
electron system: the spatial scale $\xi_{\rm pin}$ of the low-energy
excitations is much larger than the mean free path $l(T)$ which characterizes
their ``center-of-mass" diffusion. Counterintuitively, the Anderson
localization length $\xi_{\rm loc}$ vanishes to zero in the classical
limit. This implies that the conductance at low $T$ will be strongly
suppressed for $L$ larger than $\xi_{\rm loc}$, which is much smaller than
$\xi_{\rm pin}$.

We finish with a brief discussion of the $T$ dependence of $\sigma_{\rm dc}$
for $T\ll \Delta_{\rm cl}$. Since in 1d the averaging over disorder assumes
summation of random resistances connected in series, the activation over
barriers whose typical height is $\Delta_{\rm cl}$ and statistics is Gaussian
is described by $\ln\sigma_{\rm dc}\propto -(\Delta_{\rm cl}/T)^2$, rather
than by the Arrhenius law. This falloff continues with decreasing $T$ down to
the critical temperature $T_c\sim T_1$ [Eq.~(\ref{21})], below which
$\tau_\phi^{-1}(T)$ and $\sigma_{\rm dc}(T)$ vanish to zero \cite{remark4},
similarly to Ref.~\onlinecite{gornyi05b}. Since $\xi_{\rm pin}/\xi_{\rm
loc}(T)\sim |\ln\sigma_{\rm dc}(T)|$ for $\Delta_{\rm cl}\gg T\gg T_1$, the
presence of the quasiclassical barriers does not affect the estimate for
$T_c$.

To summarize, there are two distinctly different disorder-induced phenomena
that limit the mobility of CDWs: quantum localization and classical
pinning. We have discussed their interplay in the transport properties of CDWs
in 1d. The pinning turns out to be stronger in the respect that the pinning
temperature $\Delta_{\rm cl}$ is larger than the critical temperature $T_c$ of
the localization transition. On the other hand, the localization is stronger
in that the quantum localization length $\xi_{\rm loc}$ is shorter than the
pinning length $\xi_{\rm pin}$. A rigorous analytical description of the
localization of CDWs [in particular, a derivation of
Eqs.~(\ref{19}) and (\ref{20})] is clearly warranted. More generally, the
challenge is to study the dependence of the conductivity $\sigma (\omega, q,
T)$ as a function of frequency, momentum, and temperature, where both
phenomena and corresponding scales in all three variables will show up. Here,
we have addressed the $L$ dependence of the dc conductance and the $T$
dependence of the conductivity at $\omega,q=0$.

We thank A.~Glatz, L.~Glazman, O.~Yevtushenko, and especially I.~Gornyi for
discussions. This work was supported by the CFN of the DFG and by the U.S. DOE
under Contract No.\ DE-AC02-06CH11357. Two of us (A.D.M and D.G.P) 
acknowledge the
hospitality of the Materials Theory Institute at the ANL. 
%%%%%{\it Note added:}
When this work was completed, we became aware of the preprint
\cite{rosenow06}, where a similar calculation of $\sigma(\omega,T)$ was
presented. We thank B.~Rosenow for discussion and sharing with us the results
of Ref.~\onlinecite{rosenow06} prior to its publication.


\begin{thebibliography}{1}
\bibitem{remark1} 
%The notion of creep \cite{blatter94} implies that
%$\Delta_{\rm cl}$ diverges with decreasing driving force. In 1d, however, this
%phenomenon is absent.
Creep \cite{blatter94}, implying that $\Delta_{\rm cl}$ diverges with
decreasing driving force, is absent in one dimension.
\bibitem{recall} To recall the results for weak interaction ($a=1-K\ll 1$)
\cite{gornyi05a}, for spinless electrons in the Luttinger liquid
$T_1\sim (w/a^2u)(u^2/w\lambda)^{2a/(1+2a)}$ and the critical temperature
$T_c\sim aT_1/\ln(1/a)\ll T_1$. For $a^2T_1\alt T\alt T_1$, $\xi_{\rm loc}\sim
(u^2/w)(T\lambda/u)^{2a}$. These results cannot be straightforwardly
generalized to the case of strong interaction with $K\ll 1$, in particular
because the velocities $u$ and $c$ become then parametrically different from
each other.  %\cite{remark3}.
\bibitem{remark5} Although Eq.~(\ref{17}) is valid at $T\gg\Delta_{\rm cl}$,
using Eq.~(\ref{18}) for the Drude backscattering time is legitimate
down to $T\sim T_1$, i.e., as long as the spatial scale $u/T$ that determines
$\tau (T)$ (the correlation radius of Friedel oscillations induced by
disorder) is shorter than $u\tau (T)$.
\end{thebibliography}
\end{document}